# Controlling the band gap of ZnO by programmable annealing

*Shouzhi Ma,[1] Houkun Liang,[2] Xiaohui Wang,[3] Ji Zhou,[3] Longtu Li,[3] and Chang Q Sun[1,]\**

[1] School of Electric and Electronic Engineering, Nanyang Technological University, Singapore 639798

[2] Singapore Institute of Manufacturing Technology, 71 Nanyang Drive, Singapore, 638075

[3] State Key Laboratory of New Ceramics and Fine Processing, Department of Materials Science and Engineering, Tsinghua University, Beijing 100084, China

\* Corresponding Author, Phone: (65)67904517; Electronic mail: ecqsun@ntu.edu.sg

**ABSTRACT**

Annealing has been extensively used to control crystal growth and physical properties of materials with unfortunately unclear mechanism and quantitative correlations. Here we present the "annealing temperature - grain size - band gap" correlation for ZnO nanocrystals with experimental evidence. Findings revealed that the annealing condition determines the critical size by equating the thermal and the cohesive energy of the undercoordinated atoms in the surface skin, which in turn induce local strain and quantum entrapment, perturbing the Hamiltonian and hence the band gap. The formulation provides a general guideline for controlling crystal growth and performance of materials, and makes predictive design and fabrication of functional nanomaterials into reality.

KEYWORDS: size effect, photoabsorption, photoluminescence, BOLS

I. INTRODUCTION

With the miniaturization of crystal size, the fraction of undercoordinated surface atoms becomes dominant, and hence, materials in the nano-regime behave differently from their bulk counterpart. Many features of the material are no longer constant but become tunable with size. For example, the band gap ($E_G$) of ZnO nanostructures becomes tunable,[1] which is of profound importance to the optical device design such as light emitters[2] and laser diodes.[3] One effective means being widely used to tune the grain size and the $E_G$ of nanoparticles (NPs) is thermal annealing. Experimentally, the narrowing of the $E_G$ resulted from annealing ZnO NPs has been observed by Zak *et al.*[4] By comparing the photoabsorption (PA) and photoluminescence (PL) spectra, Lin *et al.*[5] observed an increase in the Stoke shift as the particle size is decreased. However, one challenging issue is that the annealing process is usually based on a trial-error approach which lacks theoretical guideline. On the other hand, the effect of particle size on the energy of photoabsorption ($E_{PA}$),[6] photoluminescence ($E_{PL}$),[7] and their correlation remain unclear despite models such as quantum confinement,[8] thermodynamics,[9] luminescence center,[10] free-exciton collision,[11] and surface states.[12] These models explain phenomenologically well the energy shift of either the $E_{PA}$ or the $E_{PL}$ alone. However, the relationship among annealing temperature, grain size, $E_{PA}$, and $E_{PL}$ remains a challenge. A quantitative correlation between the annealing temperature and the derived physical properties pursued is therefore highly desired. In fact, the annealing temperature, the crystal size, and the band gap are interconnected. In this work, we propose and verify these size related properties from the perspective of bond order-length-strength (BOLS) correlation mechanism[13] with ZnO nanocrystals as the prototype of demonstration.

## II. THEORY

According to the BOLS theory,[13] the size tenability at nanoscale arises from the tunable fraction of the undercoordinated atoms and the interaction between them.[14] Recent progress showed that not only the bond length but also the bond energy changes at the surface skin of NPs in comparison with the bulk.[15] The shorter and stronger bonds between the undercoordinated atoms[16] provide perturbation to the Hamiltonian of the entire system especially when the system is miniaturized. As a result, the physically detectable quantities undergo a modification $Q(R_{NP})=Q(\infty)(1+\Delta)$, where $Q$ denotes the physical quantity such as band gap or melting temperature, $\infty$ and $R_{NP}$ represents the bulk and finite sized NPs or grains with radius $R_{NP}$, respectively. $\Delta$ is the size-induced perturbation to $Q$. If a nanoparticle has a core-shell structure,[17] $\Delta$ solely arises from the surface atoms. The effect of undercoordinated surface atoms can be modeled by BOLS correlation: denote $d$ and $E$ as the bond length and binding energy, respectively, $d_b$, $d_i$, $E_b$, $E_i$ will be the bond length in the bulk, in the $i^{th}$ atomic layer (counting from outermost inwards), and binding energy in the bulk, in the $i^{th}$ atomic layer, respectively. For an undercoordinated atom in the $i^{th}$ surface layer, bonds contract from the ideal bulk value of $d_b$ to $d_i = c_i d_b$, and the cohesive energy per bond increases from $E_b$ to $E_i = c_i^{-m} E_b$. $c_i = 2/\{1+\exp[(12-z_i)/(8z_i)]\}$ is the bond contraction coefficient, with $z$ being the coordination number (CN) of a specific atom and $m = 4$ is the bond nature indicator for ZnO.[18] The shell consists of up to three atomic layers (i.e. $i \leq 3$), which suffers from bond order deficiency and hence bond length contraction and bond energy gain, originating the property change. Figure 1 shows the core-shell structure of a grain. When the solid size is increased, the surface to volume ratio is significantly reduced. By employing the BOLS induced perturbation, the unusual mechanical, thermal, electrical, optical, and acoustic behaviors resulted from vacancies, adatoms, defects, nanostructures and size effect can be unified.

According to the nearly-free electron approximation, $E_G$ depends uniquely on the first Fourier coefficient of the crystal potential, and is proportional to the binding energy per bond: $E_G = 2|V_1| \propto \langle E_b \rangle$. Base on the BOLS correlation, the perturbation to $E_G$ is derived as:

$$\begin{cases} E_G(R) = E_G(\infty)(1 + \Delta_H) \\ \Delta_H = R^{-1} \sum_{i \leq 3} \tau c_i \left( c_i^{-m} - 1 \right) = R^{-1} \Delta_H' \end{cases}, \qquad (1)$$

where $R = R_{NP}/d$ is the normalized radius of a nanoparticle or grain, meaning the number of atoms lined along the radius (as illustrated by Figure 1). $\Delta_H$ is the Hamiltonian perturbation and $\tau c_i R^{-1}$ is the surface-to-volume ratio, which means the proportion of atoms in the $i^{th}$ atomic layer to the total number of atoms in a NP as derived in Ref.[18] $\tau$ is the dimensionality, which equals to 2 for nanowire and 3 for NPs.

Besides the band gap, the grain radius $R$ is also tuned by annealing as presented by Figure 1. The cohesive energy, $E_{coh}$, determines the NP size allowed at certain temperature as the NP size is directly related to the cohesive energy.[19] We have demonstrated that the size of a particle in nucleation depends on the ratio between the annealing ($T_a$) and the melting temperature ($T_m$) of the specimen, $T_a/T_m$.[20] When the $T_a \approx 0.3 T_m$, the particle growth is in equilibrium and the size is stable.[21] Larger grains require higher $T_a$,[22] but as a competing factor, the $T_m$ increases with grain size.[23] It is essential to find the size effect on the $T_m$ before quantifying the $R$ to be tuned by $T_a$.

$T_m$ is proportional to the atomic cohesive energy, and equals to the product of bond energy and the atomic CN, i.e. $E_{coh} = zE_b$, which results in the cohesive energy perturbation $\Delta_{coh}$ containing $z$:

$$\begin{cases} T_m(R) = T_m(\infty)(1 + \Delta_{coh}) \\ \Delta_{coh} = R^{-1} \sum_{i \leq 3} \tau c_i \left( z_i/z_b \times c_i^{-m} - 1 \right) = R^{-1} \Delta'_{coh} \end{cases} \qquad (2)$$

For the post-annealing process, the as-grown particle size ($R_0$) and threshold temperature ($T_{th}$) need to be involved. The high-energy grain boundary does not gain mobility until reaching $T_{th}$,[24] at which grains grow upon heating to minimize the overall energy. With $T_{th}$ and $R_0$ in consideration, Eq. (3) gives the completed expression for critical size $R$ allowed by the respective $T_a$:

$$\begin{cases} T_a - T_{th} = 0.3 T_m(R) = 0.3 T_m(\infty)\left(1 + R^{-1}\Delta'_{coh}\right) \\ R - R_0 = \dfrac{\Delta'_{coh}}{(T_a - T_{th})/[0.3 T_m(\infty)] - 1} = \dfrac{|\Delta'_{coh}|}{1 - (T_a - T_{th})/[0.3 T_m(\infty)]} \end{cases} \qquad (3)$$

This relation indicates that the crystal size is dominated by the term $(T_a - T_{th})/[0.3T_m(\infty)]$; grain grows as $T_a$ rises when $T_a > T_{th}$. For a given $T_a$, the grain radius $R$ is hence predictable.

III. EXPERIMENTS

In order to verify the theoretical predictions of grain size and band gap tuned by the annealing process, polycrystalline ZnO thin film of 10 nm grain radius was deposited on quartz substrate using an arc deposition technique.[25] A post-annealing process at 573 - 973 K in air for 1 hour tuned the grain radius to 10 - 40 nm. The grain sizes were measured using the X-ray diffraction (XRD) and scanning electron microscope (SEM). XRD measurements revealed that the samples were polycrystalline with a hexagonal wurtzite structure, and had a preferred (002) orientation. In the ZnO films, the amount of impurities and defects is neglectably small. The grain size was determined from the full width at half maximum (FWHM) of the (002) peak (i.e. 2θ = 34.5°) of the XRD profiles, according to Scherrer's equation with λ = 0.154 nm for CuK$_α$ radiation. Table 1 summarizes the grain size derived from XRD and SEM results. The SEM inspection shows agreement with the XRD measurements. To study the size effect on the optical band gap at room temperature, PL spectrum was acquired using the Nd:YAG (355 nm) pulsed laser (10 Hz). The absorption was measured on a double-beam spectrophotometer. The measured $E_{PA}$ and $E_{PL}$ for different grain sizes are also summarized in Table 1.

IV. RESULTS AND DISCUSSIONS

The as-grown grain radius is 10 nm, so $R_0$ = 10 nm / 0.199 nm[1] = 50 after normalization. There is no obvious grain growth below 573 K, indicating that 573 K is the $T_{th}$. According to Eq. (3), grain size tuned by different $T_a$ can be theoretically predicted as shown in Figure 2. Experimental results show good agreement with expectations, especially before the grains gain fast growth. The stable grain size increases rapidly with temperature when the $T_a$ is above 900 K. At higher $T_a$, although there is an increase in discrepancy between experiments and calculation, the deviation does not have strong effect as the grains are in the sub-micron regime and surface atoms gradually lose their hegemony. The inset shows the SEM micrograph of spontaneous grain growth at different temperatures. By employing the coordination and bond length correlation, the cohesive energy is found to be proportional to the inverse grain radius. As $T_a$ increases, the stable grain size is determined by the decreased surface to volume ratio. Consistency between experiment and calculation confirms the validity of BOLS theory, which can provide guidelines for controlling thermal annealing.

The absorption coefficient was derived from the transmission spectra according to $I = I_0\exp(-\alpha x)$. For direct band gap materials, the absorption coefficient $\alpha \propto \sqrt{(h\nu-E_G)}$. $\alpha^2$ versus photon energy $h\nu$ was derived from the transmission spectra as shown in Figure 3(a). By using the linear fitting of the absorption edges, the $E_{PA}$ were derived and found to be blue shifted with the decrease of the particle size. To estimate the difference between $E_{PA}$ and $E_{PL}$, the PL spectra were also shown in Figure 3(a) for comparison. In the PL spectra, near-band-edge emissions were observed in all of the films. The peak positions were also blue shifted with the decrease of the particle size. The shift arises from the size effect as illustrated by Eq. (1). Comparing the PA and PL spectra, one can note that the $E_{PL}$ is lower than the $E_{PA}$. The energy offset, i.e. Stoke shift ($E_{SS}$), as theoretically[26] explained and experimentally[27] verified, is due to the lattice vibration and exciton binding energy. In our experiment, $E_{SS}$ is around 0.1 eV, which agrees with documented values[28] except that Lin et al.[5] has observed a much larger $E_{SS}$. Since $E_{SS}$ comes from electron-phonon coupling and crystal binding[29] in a crystal lattice, it is size dependent: for a single bond, $E_{SS} \propto q^2$ with $q$ being in the dimension of wave vector. $q$ is inversely proportional to atomic distance $d$, and hence, $E_{SS} \propto A/d^2$, in the surface region where $A$ is a constant depending on the band curvature. It has been already shown that the bonds shrink in the surface layers. Based on the BOLS premise, $E_{SS}$ can be correlated to the bond contraction and expressed as:

$$E_{SS} = AR^{-1}\sum_{i\leq 3}\tau c_i\left(c_i^{-2}-1\right)E_{PL} + 0.1. \tag{4}$$

Eq. (4) indicates that $E_{SS}$ consists of two terms: first term represents the size-originated energy shift due to bond contraction, and the second term is the empirical bulk $E_{SS}$ which is around 0.1 eV.[28] In order to estimate the size effect on $E_{PL}$ and $E_{PA}$, both the Hamiltonian perturbation and the Stoke shift should be involved. Figure 3(b) shows the comparison of the experimentally observed and theoretically predicted energy shift $\Delta E$ in PA and PL spectrum of finite sized NPs with respect to their corresponding bulk energy $E(\infty)$ versus normalized grain radius. The calculated $E_{PA}$ and $E_{PL}$ are $E_{PA} = E_G + 1/2E_{SS}$ and $E_{PL} = E_G - 1/2E_{SS}$, where $E_G$ and $E_{SS}$ are given in Eq.(1) and Eq.(4) respectively. In the process of excitation and recombination, an electron absorbs a photon with energy $E_{PA} = E_{PL} + E_{SS}$ and emits a photon with energy $E_{PL}$. At the surface, the CN-imperfection-enhanced bond strength affects both the Stoke shift and the band gap. Good agreement between predictions and measurements proved the validity of the BOLS correlation. Our experiments suggested that $E_{PA}$ and $E_{PL}$ for bulk are 3.15 and 3.05 eV, respectively. Figure 4 shows the good match of the calculated $E_{SS}$ based on Eq. (4) with experimental data. The large $E_{SS}$ in Lin's experiment (i.e. $R < 40$) can be explained by size effect. The grain size is so small that the size-dependent term in Eq.(4) dominates; the first term in Eq.(4) is significant enough to increase the $E_{SS}$ from 0.1 eV to 0.3 eV due to increased surface to volume ratio. The blue shift of $E_{PA}$ and $E_{PL}$ is the joint contribution from the crystal binding and the electron-phonon coupling. Without the bond contraction, neither $E_G$ expansion nor PA or PL blue shift will happen; the summation over all the volume will not be $R$ dependent.

There are a number of experiments on optical properties of ZnO nanostructures which supports our analytical model. For example, Zak et al.[4] found that annealing increases the size for ZnO NPs, associated with a narrowing of the optical band gap. The PL and PA blue shift and Stoke shift of ZnO quantum dots measured by Lin et al.[5] are in agreement with the BOLS correlation. The same trend of Stoke shift has also been observed for other direct band gap materials besides ZnO.[30] Documented experimental band gap varies with synthesis methods and measuring technique. In our experiment, the bulk absorption band gap is 3.15 eV, which is smaller than the literature.[1] Srikant and Clarke's[31] observed that the band gap of ZnO derived from absorption was lower than that from other methods such as spectroscopic ellipsometry, possibly because the transmittance measurement probe both the surface and bulk of the sample while other popular methods only the surface. This conclusion provides an evidence for the significance of bond deficiency and bond enhancement at surfaces.

V. CONCLUSION

In conclusion, an analytical correlation between the "annealing temperature - grain size - band gap" has been established towards consistent understanding of the annealing controlled ZnO synthesis and band gap. The findings demonstrated that the annealing induced grain size and band gap change are attributed to the surface and interface atomic CN deficiency and bond energy perturbation. Annealing tunes the equilibrium atomic cohesive energy while the perturbation to Hamiltonian determines the band gap. Exceedingly good agreement between predictions and measurements indicates that the broken bond in the surface layers is the essential cause for the temperature driven grain growth and spontaneous band gap shift.

ACKNOWLEDGMENT The work was supported by National Science Fund for distinguished young scholars (No. 50625204), Science Fund for Creative Research Groups (No. 50621201), and by the Ministry of Sciences and Technology of China through National Basic Research Program of China (973 Programs under Grant No. 2009CB623301).

FIGURE CAPTIONS

Figure 1 Schematic of the core-shell structure of a single grain, and grain growth, band gap evolution resulted from thermal annealing.

Figure 2 Comparison of theoretically predicted $R$ (normalized) with experimentally measured from XRD profile. The inset shows the SEM micrograph of ZnO NPs annealed at different temperatures: (i) as-grown grains, and grains annealed at (ii) 773 K, (iii) 873 K, and (iv) 923 K.

Figure 3 (a) Experimentally measured photoluminescence and absorption spectrum for various grain diameters. (b) Experimental and theoretical $E_{PA}$ and $E_{PL}$ blue shift for different grain radius $R$ (normalized).

Figure 4 Experimentally measured and theoretically calculated size dependent Stoke shift for various grain radius $R$ (normalized). Data of Exp. (1) is from Ref,[5] and Exp. (2) is from our experiment.

TABLE 1 Grain radius derived from XRD FWHM and SEM micrographs for ZnO annealed at different temperatures, and experimentally measured $E_{PA}$ and $E_{PL}$.

| Temperature (K) | FWHM | XRD (nm) | SEM (nm) | $E_{PA}$ (eV) | $E_{PL}$ (eV) |
|---|---|---|---|---|---|
| 573 | 0.4 | 10 | 10 | 3.339 | 3.234 |
| 673 | 0.37 | 11 | 11 | 3.330 | 3.213 |
| 773 | 0.29 | 14.5 | 15 | 3.315 | 3.202 |
| 823 | 0.25 | 16.5 | 17.5 | 3.282 | 3.142 |
| 873 | 0.22 | 18.5 | 18.5 | 3.264 | 3.123 |
| 923 | 0.21 | 20 | 25 | 3.249 | 3.108 |
| 973 | 0.16 | 25 | 40 | 3.220 | 3.099 |

SYNOPSIS GRAPH:

Equilibration of the thermal and the cohesive energy of the undercoordinated skin atoms determines the grain size and hence the size-dependent band gap due to the skin-region quantum entrapment.

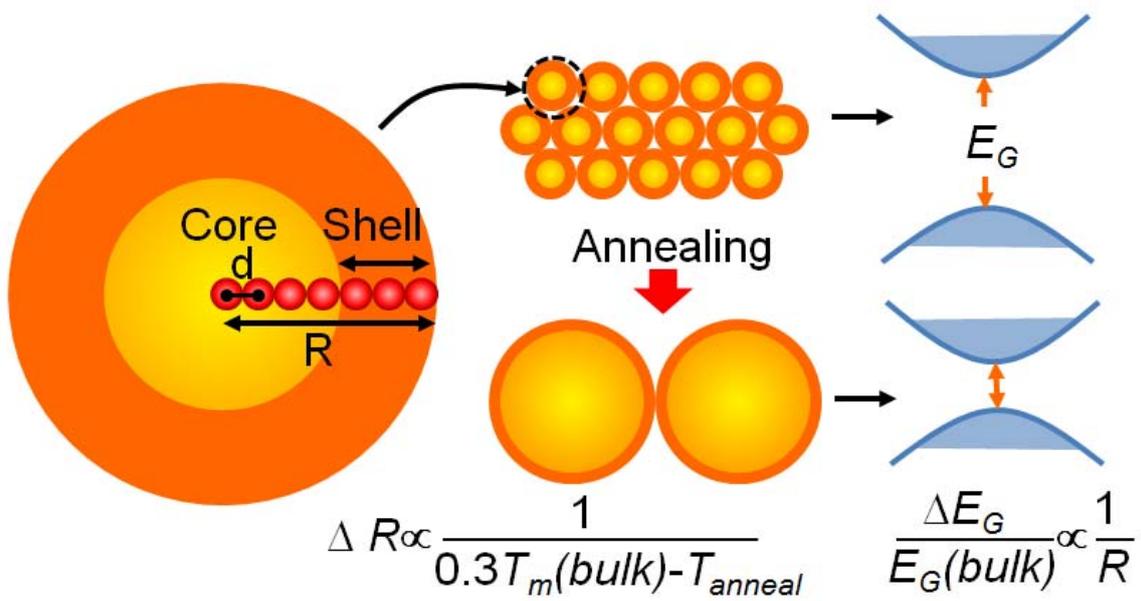

Figures
Figure 1

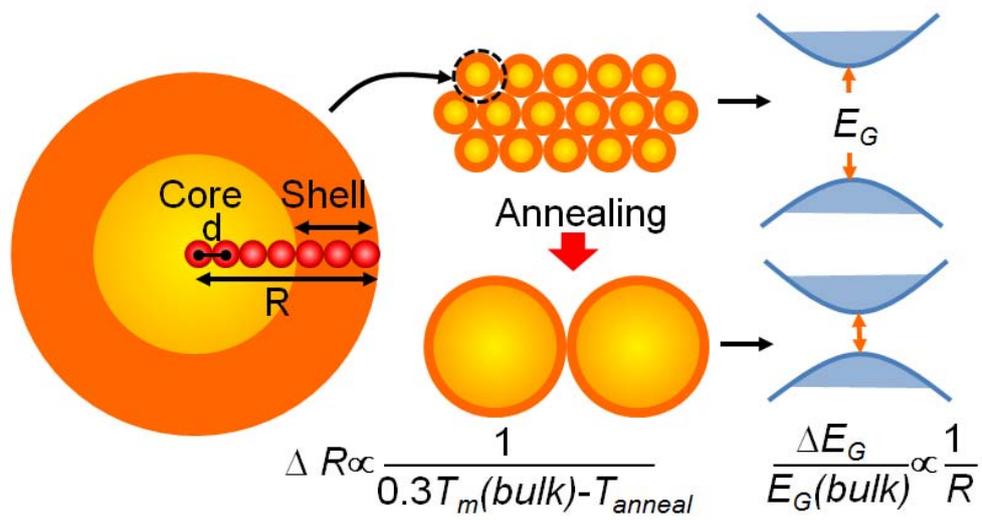

Figure 2

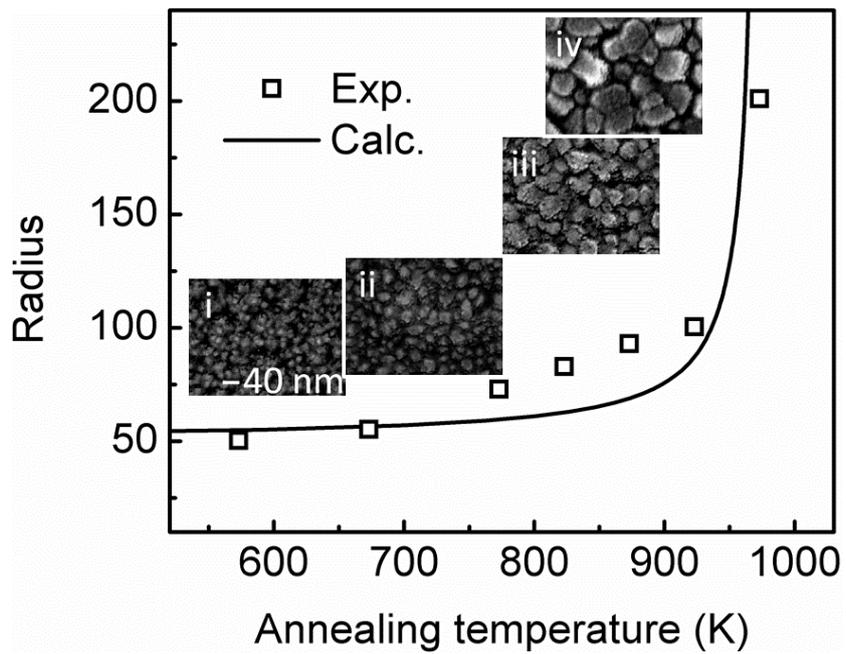

Figure 3

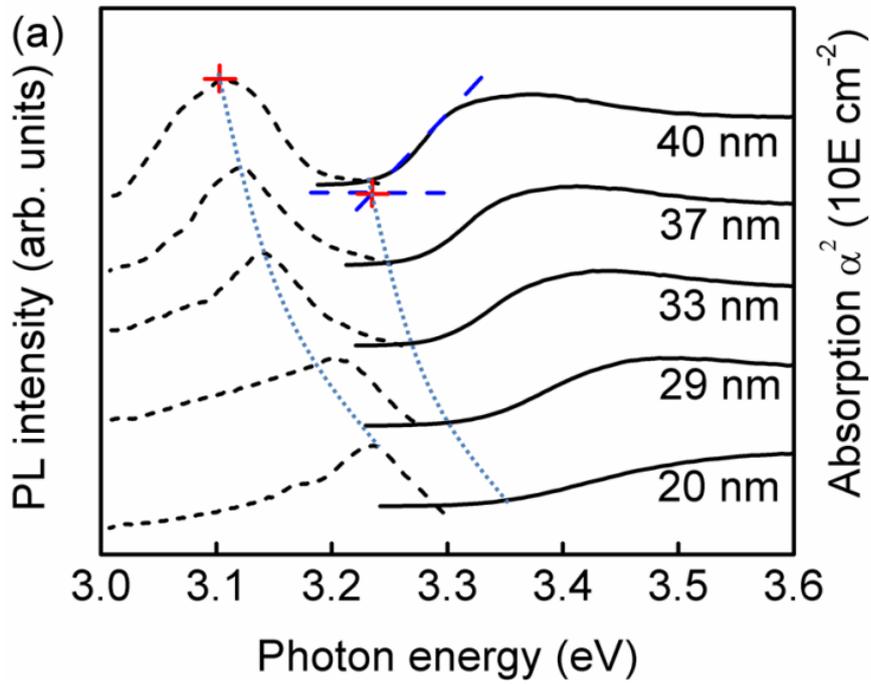

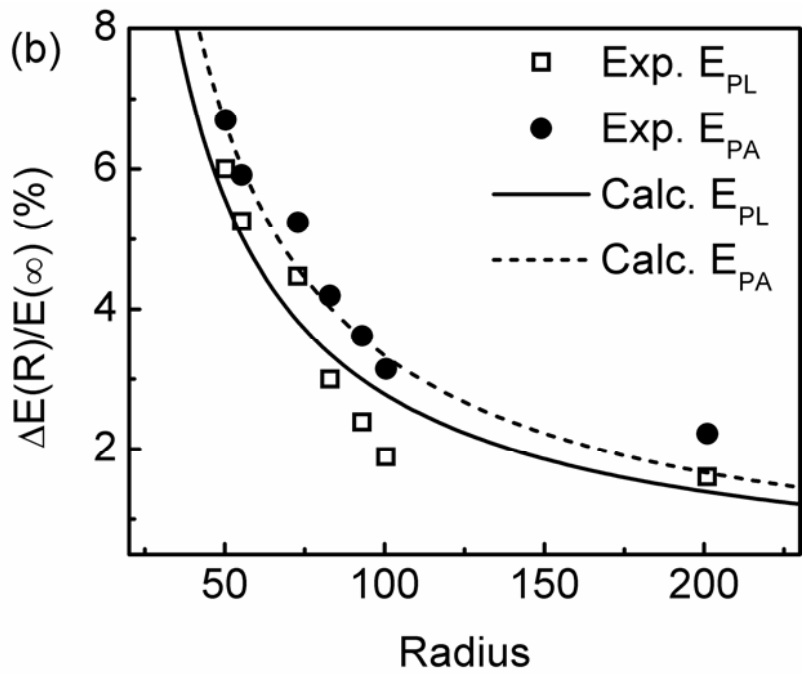

Figure 4

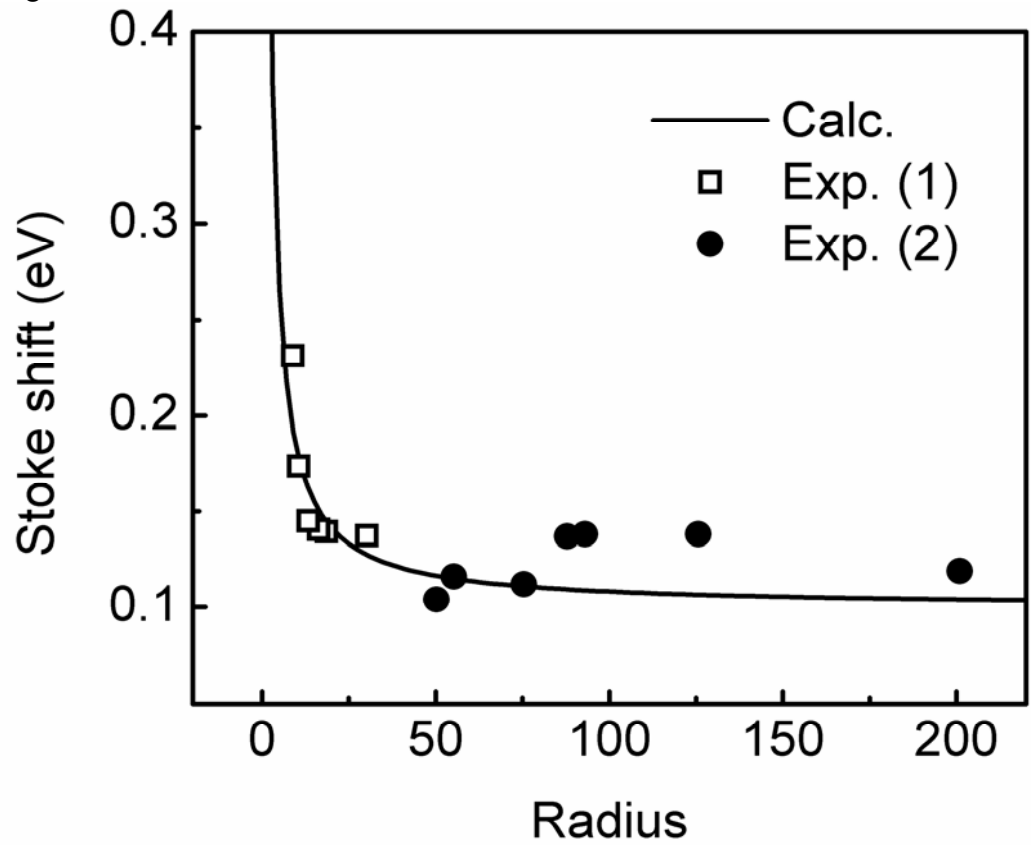